\documentclass[%
 aip,
 amsmath,amssymb,
 reprint,%
]{revtex4-1}

\usepackage{graphicx}
\usepackage{caption,subcaption}
\usepackage{dcolumn}
\usepackage{bm}

\usepackage[utf8]{inputenc}
\usepackage[T1]{fontenc}
\usepackage{mathptmx}
\usepackage{xcolor}
\usepackage[normalem]{ulem}

\begin{document}

\preprint{AIP/123-QED}

\title[MIT-Pisa title]{Magnetic Induction Imaging with a cold-atom radio-frequency magnetometer}

\author{A. Fregosi}
\affiliation{Istituto Nazionale di Ottica, CNR-INO, via Moruzzi 1, 56124 Pisa, Italy}%
\affiliation{Department of Physical Sciences, Earth and Environment, University of Siena, 
via Roma 56, 53100 Siena, Italy}
\author{C. Gabbanini}
\affiliation{Istituto Nazionale di Ottica, CNR-INO, via Moruzzi 1, 56124 Pisa, Italy}
\author{S. Gozzini}
\affiliation{Istituto Nazionale di Ottica, CNR-INO, via Moruzzi 1, 56124 Pisa, Italy}
\author{L. Lenci}
\affiliation{Facultad de Ingenier\'ia, Instituto de F\'isica, Universidad de la Rep\'ublica, J. Herrera y Reissig 565, 11300 Montevideo, Uruguay}
\author{C. Marinelli}
\affiliation{Department of Physical Sciences, Earth and Environment, University of Siena, 
via Roma 56, 53100 Siena, Italy}
\affiliation{Istituto Nazionale di Ottica, CNR-INO, via Moruzzi 1, 56124 Pisa, Italy}
\author{A. Fioretti}%
\email{andrea.fioretti@ino.cnr.it}
\affiliation{Istituto Nazionale di Ottica, CNR-INO, via Moruzzi 1, 56124 Pisa, Italy}
%


\date{\today}

\begin{abstract}
The sensitive detection of either static or radio-frequency \textsc{(rf)} magnetic fields is essential to many fundamental studies and applications. Here, we demonstrate the operation of a cold-atom-based, \textsc{rf} magnetometer in performing {1-D and 2-D imaging} of small metallic objects. It is based on a cold $^{85}$Rb atomic sample, and operates in an unshielded environment with no active field stabilization. It  shows a sensitivity up to $200\, \rm{pT}/\sqrt{\rm Hz}$ in the $5-35\, \rm{kHz}$ range bandwidth and can resolve a cut 
$0.4\, \rm{mm}$ wide in a $0.8\, \rm{mm}$ thick metallic foil. The characteristics of our system make it a good candidate for applications in civil and industrial surveillance. 
\end{abstract}

\maketitle


\label{sec:intro}

The development of magnetometers and magnetic sensors has been a central goal for many decades for its strong implications in fundamental research and applications. Many types of magnetometers with very different sensitivity, range, dimensions and cost, are available in various domains: geology, space research, biology, medicine and civil, industrial and military security~\cite{Ripka2001}.

In high-sensitivity applications, optical atomic magnetometers~\cite{Budker2013} (OAMs) compete with superconducting quantum interference devices~\cite{Fagaly2006} in attaining record sensitivity (well below one fT/$\sqrt{\rm Hz}$), spatial resolution and measurement bandwidth. OAMs are based on the optical detection of the effect of a {static}  or \textsc{rf} magnetic field sensed by an optically pumped atomic medium, either in vapor phase~\cite{Budker2007} or embedded in a solid matrix~\cite{Rondin2014}. 
\textsc{rf} atomic magnetometers~\cite{Savukov2005} detect weak magnetic fields in the hundreds of Hz to hundreds of MHz frequency range, in unshielded environments.
They were applied to nuclear quadrupole resonance and low-field {nuclear magnetic resonance}~\cite{Lee2006, Savukov2009, Bevilacqua2009, Bevilacqua2019}, magneto-cardiography~\cite{Belfi2007, Alem2015} and magnetic induction {imaging or} tomography~\cite{Wickenbrock2014, Deans2016, Wickenbrock2016, Deans2017, Deans2018, Deans2018b, Bevington2019, Deans2020}. 
{The latter} technique is a tool for diagnostic and surveillance, capable of 3-D imaging of magnetic/conducting objects and their defects~\cite{Griffiths2001, Wei2013}. 

Although OAMs employing atomic vapors are very powerful, they suffer limitations in terms of spatial resolution and interrogation times in reason of the atomic spin diffusion and decoherence. Beyond the bright solutions adopted so far to mitigate these problems in thermal vapors, like the introduction of an inert buffer gas~\cite{Kominis2003}, %
 the anti-relaxation surface coating~\cite{Seltzer2009} and the miniaturization~\cite{Shah2007} 
of glass cells, another useful approach is that of using ultracold atoms~\cite{Metcalf1999}.

Cold atoms in optical molasses~\cite{Isayama1999}, dipole traps~\cite{Koschorreck2011} and Bose-Einstein condensates~\cite{Vengalattore2007} are very appealing for magnetometry because, despite their relatively low number as compared to a thermal sample, they provide high spatial resolution and long coherence times.
A cold-atom-based \textsc{rf} magnetometer (\textsc{rf} C-AM) has been recently realized~\cite{Cohen2019} to operate in an unshielded environment, showing a sensitivity up to $330\, {\rm pT}/\sqrt{\rm Hz}$. This important result is somehow limited by its slow repetition rate, of the order of $0.1\, {\rm Hz}$, that prevents extensive measurements in stable experimental conditions.

In this Letter we show the possibility of {producing conductivity maps} of metallic objects with a  \textsc{rf} C-AM {as a function of the  source-object relative position}. This has been possible thanks to the large improvement in the repetition rate of the measurements,  up to $5\, {\rm Hz}$ or more. 
We show 1-D {profiles} and 2-D {images} of  metallic objects down to a 
${2.25}\, {\rm cm}^2$ surface and a $0.8\, {\rm mm}$ thickness, and detect cuts as narrow as $0.4\, {\rm mm}$. Images are taken in an unshielded environment, without active background magnetic field control and without the subtraction of a background image. The sensitivity of the present setup is already below $1\, {\rm nT}/\sqrt{\mathrm{Hz}}$ in its frequency  range ($1-50\, {\rm kHz}$), representing  a step forward in the direction of applying cold-atom-based magnetic sensors for civil, industrial and military monitoring.

\label{sec:setup}

\begin{figure*}[htb]
\centering
%
%
  \includegraphics[width=0.9\linewidth]{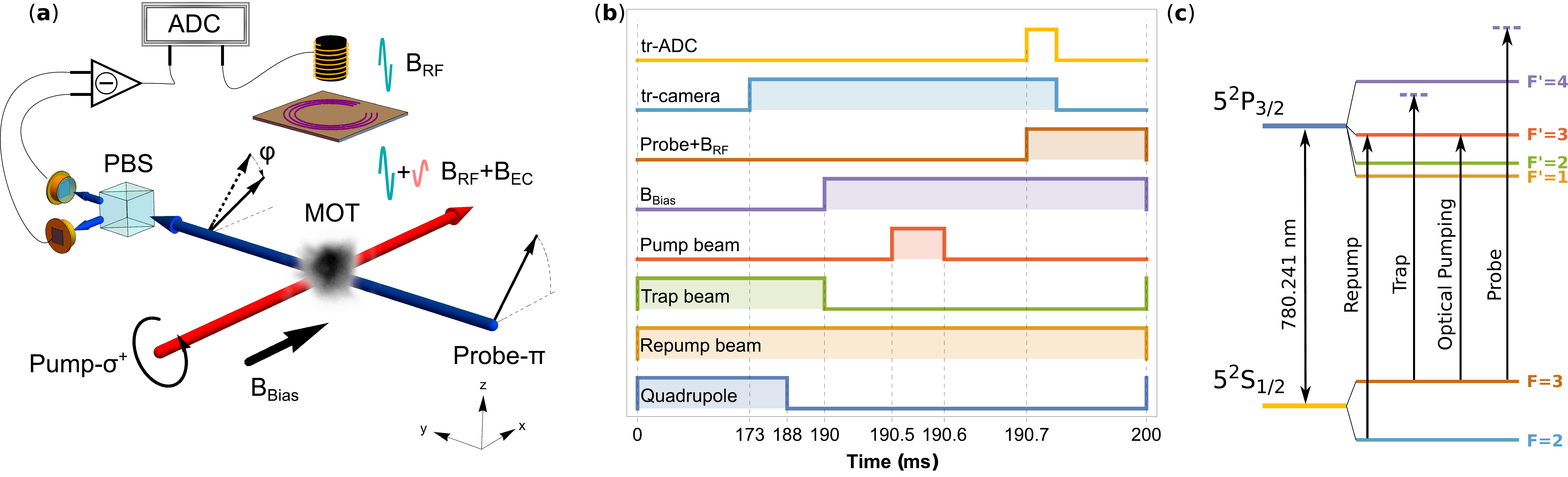}
%
%
\caption{In a) sketch of the working principle of the \textsc{rf} C-AM; in b) temporal sequence of the experiment, described in the text (notice that the scale is not linear); in c) relevant level scheme and optical transitions for the magneto-optical trapping, optical pumping and probing of the atoms.}

\label{fig:setup}
\end{figure*}

The \textsc{rf} C-AM is composed of a standard setup to realize a magneto-optical trap (MOT) for Rb atoms, and of the apparatus to {produce conductivity maps} of small metallic objects. The principle of the measurements (Fig.~\ref{fig:setup}{a}) is as follows: a sample of Rb atoms is trapped, laser-cooled to tens microKelvin, released and spin-polarized in a uniform magnetic field, ${\bf B}_{\rm Bias}$. A \textsc{rf} magnetic field, ${\bf B}_{\textsc{rf}}$, orthogonal to the atomic polarization, induces an atomic magnetization that starts rotating at the Larmor frequency. A linearly polarized laser beam, impinging on the atoms along the third orthogonal direction, probes this magnetization through the detection of the Faraday rotation. In presence of a metallic object, the eddy currents generate an additional field, ${\bf B}_{\textsc{ec}}$, which modifies the amplitude and the phase of the total field on the atoms, and thus the magnitude of the Faraday rotation. The {conductivity maps are obtained} by recording, through a lock-in amplifier, amplitude and phase of the signal of a polarimeter, which analyzes the probe polarization, as a function of the object position. 
As the MOT uses resonant cw lasers and a magnetic field gradient~\cite{Metcalf1999} that are not compatible with the OAM operation, the experiment works in cycles (Fig.~\ref{fig:setup}{b}) with a typical repetition rate of $5\, {\rm Hz}$. The magnetometer signal is recorded during $10\, {\rm ms}$ when all MOT lasers\footnote{{In reality, one of the two MOT lasers, the repumper, is left on during the whole cycle, because in this condition we observed a larger magnetometer signal.}} and fields are off and the slowly expanding atoms are interrogated by the probe, with a duty-cycle of 1/20. 


The MOT is produced, under {ultra-high vacuum} conditions, inside a small rectangular pyrex cell ($30\times30\times80\,{\rm mm}^3$) by three pairs of counter-propagating, orthogonal laser beams and a quadrupolar magnetic field with a gradient of about $20\,{\rm G/cm}$. An optimal background pressure of Rb vapor is obtained by setting an appropriate current through a Rb dispenser. 
All laser beams necessary for the experiment (trapping, repumping, optical pumping and probe lasers) are produced by three External Cavity Diode Lasers (ECDLs), frequency-stabilized  using saturated absorption spectroscopy and controlled in intensity and timing by Acousto-Optic Modulators (AOMs) and mechanical shutters. Their frequencies, together with the $^{85}$Rb relevant atomic levels are shown in Fig.~\ref{fig:setup}{c}.
Two ECDLs are used as trapping (waist of $13\, \rm{mm}$ and intensity  $I\sim 1.6\, {\rm mW}/{\rm cm}^2$ per beam) and repumping (power $2.4\, \rm{mW}$) lasers, the latter prevents atoms to accumulate into the $F_g=2$ level and exit the cooling/trapping {or optical pumping}  process{es}. 
The circularly-polarized, optical pumping beam is derived from the trap laser and tuned to the $F_g=3 \rightarrow F_e=3$ hyperfine transition by two AOMs. The linearly-polarized probe beam is provided by a third ECDL referenced to the $F_g=3 \rightarrow F_e=4$  transition and blue-detuned by $+320 {\rm MHz}$ using two  AOMs. Its typical power is below $20\,\mu{\rm W}$ in an elliptical waist of $0.7\times1.3\,\mathrm{mm}^2$.
The polarimeter, composed of a half-wave plate, a {Polarizing Beam Splitter} (PBS) cube and a balanced pair of photodiodes in a differential configuration, detects changes in the polarization direction of the probe laser.
The ambient magnetic field is passively compensated by three pairs of square coils in almost Helmholtz configuration. A fourth pair of coils provides the bias field,  setting the operating Larmor frequency. Finally, a coil placed  above the MOT, produces the \textsc{rf} magnetic field.
The signal of the polarimeter is acquired with a commercial {Analog-To-Digital Converter} (ADC) and a Labview program, performing as a dual-phase lock-in amplifier.  The maximum \textsc{rf} frequency in the experiment is limited by the sampling rate of the ADC ($250\, \mathrm{kHz}$ in total).


\begin{figure}[htb]
	\includegraphics[width=0.4\textwidth]{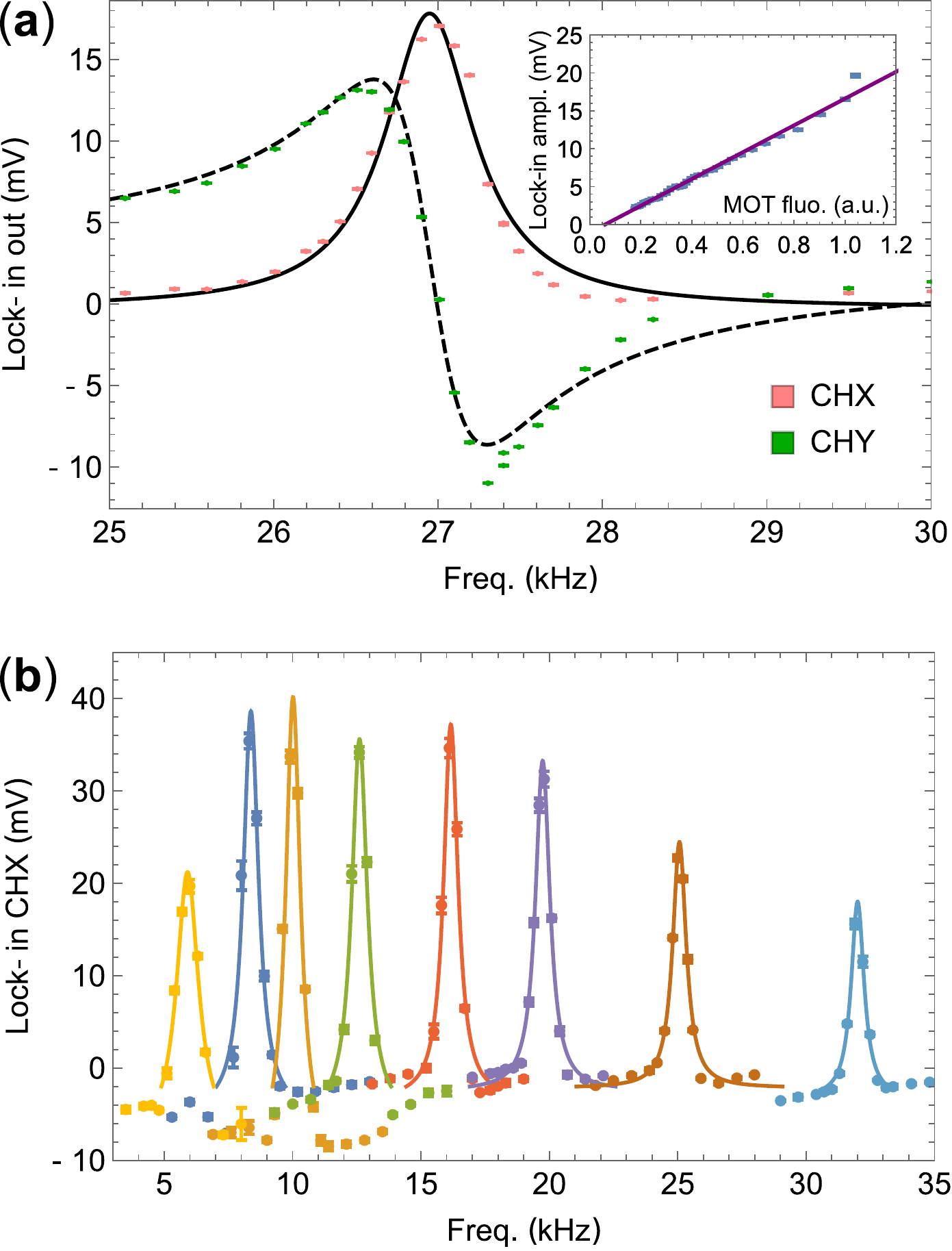}


\caption{(a) in-phase and in-quadrature response of the magnetometer fitted with a Lorentzian function (full line, FWHM $660\,\rm{Hz}$) and to a dispersive profile, respectively. The inset shows the dependence of the amplitude peak on the MOT atom number. (b) The in-phase component at different resonance frequencies.
}
\label{fig:lock-in-signal}
\end{figure}

Fig.~\ref{fig:setup}{b} shows the cycle of the experiment.
For the first $180\, \rm{ms}$ Rb atoms are loaded into the trap (MOT lasers and quadrupole field switched on). Then  atoms are further cooled, for $2\, \rm{ms}$, in an optical molasses (quadrupole field off). 
At this point ({trap} laser off, {repump laser on,} bias magnetic field $\mathbf{B}_{\rm Bias}$ on), {the combined action of} a $\sigma^{+}$ {pump} laser pulse {and of the repumper} prepares atomic spins in the stretched state $(F_g=3, m_{F}=+3)$. Finally, the \textsc{rf} field 
$\mathbf{B}_\textsc{rf}$ and the probe beam are switched on, the atoms start precessing  inducing a Faraday rotation of the probe polarization, monitored through the polarimeter for $10\, \rm{ms}$.
A {digital} CMOS camera, triggered just before the end of the loading phase, detects the MOT fluorescence giving atom number and shape of the trap (typical MOT parameters:  $5\times10^7$ trapped atoms,  $1/e$ diameter  $d\simeq0.7\,\mathrm{mm}$ and peak density $1.3\times10^{10}\,\,\mathrm{cm}^{-3}$). We observed a linear dependence of  the lock-in signals on the number of trapped atoms, as shown in the inset of Fig.~\ref{fig:lock-in-signal}a.
Magnetometer signals last up to about $10\, {\rm ms}$ because of the  ballistic expansion of the atoms  and  other decoherence processes of the atomic spins.
A careful blocking of the trapping laser with a mechanical shutter was essential to have the best signals, as well as an optimal compensation of the ambient magnetic field and field gradients.

\begin{figure}[htb]
\includegraphics[width=0.4\textwidth]{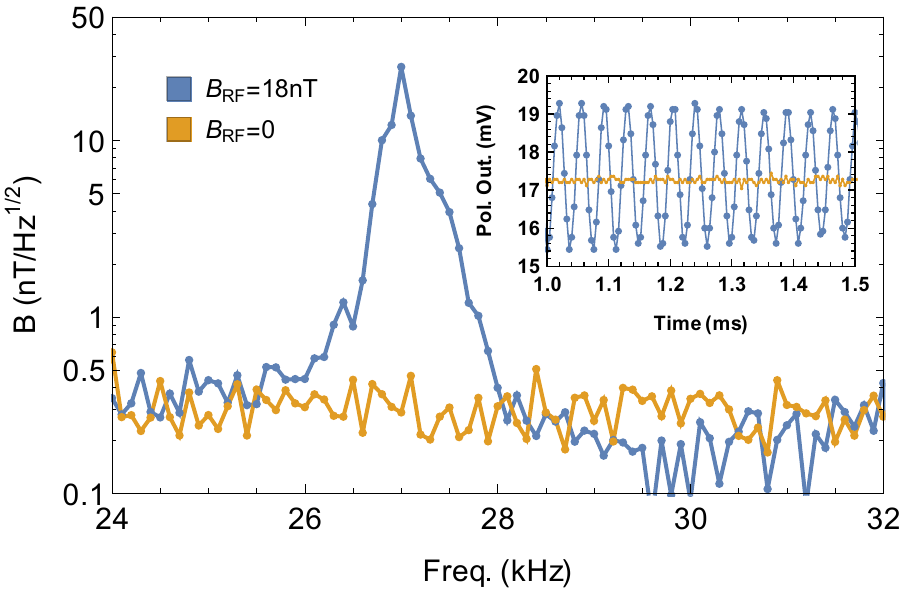}
\caption{FFT of the magnetometer response with and without a \textsc{rf} excitation corresponding to a field amplitude of $18\,\mathrm{nT}$ at the MOT position. The signal-to-noise ratio at the peak of the resonance ($f_0=27\,\mathrm{kHz}$) is $\mathrm{SNR}=86$. In the inset, a sample of the polarimeter signal with and without \textsc{rf} field.}

\label{fig:snr}
\end{figure}

A lock-in amplifier detects  the in-phase (CHX) and in-quadrature (CHY) component of the Faraday rotation with respect to the \textsc{rf} field. CHX as a function of the \textsc{rf} frequency (Fig.~\ref{fig:lock-in-signal}a) is described by a Lorentzian curve peaked at 
$\omega=\gamma {B}_\mathrm{x}$ where $\gamma=0.47\,\mathrm{kHz/mG}$ is the gyromagnetic factor for the ground state of the $^{85}\mathrm{Rb}$~\cite{Steck2008}. The linewidth of the in-phase response is roughly constant  in the $5-35 \,\mathrm{kHz}$ frequency range (Fig.~\ref{fig:lock-in-signal}b) and of the order of $600 \,\mathrm{Hz}$.


We determine the sensitivity to the \textsc{rf} field at resonance by~\cite{Cohen2019}
\begin{equation}
\label{eq:sensibilitarf}
\delta{B}_\textsc{rf}=\frac{{B}_\textsc{rf}}{SNR}
\end{equation}
where $B_\textsc{rf}$ is calibrated at the MOT position and the $SNR$ is calculated by the ratio of the {Fast Fourier Transform} (\textsc{FFT}) of the polarimeter signal with and without $\textsc{rf}$ field 
(Fig.~\ref{fig:snr}). We obtained the maximum sensitivity $\delta B_\textsc{rf}=200\,\mathrm{pT}/\!\sqrt{\mathrm{Hz}}$ in the $5-35 \,\mathrm{kHz}$ frequency range, and below $1\,\mathrm{nT}/\!\sqrt{\mathrm{Hz}}$ in the explored $1-50 \,\mathrm{kHz}$ range.

\begin{figure*}[htb]
\includegraphics[width=0.9\textwidth]{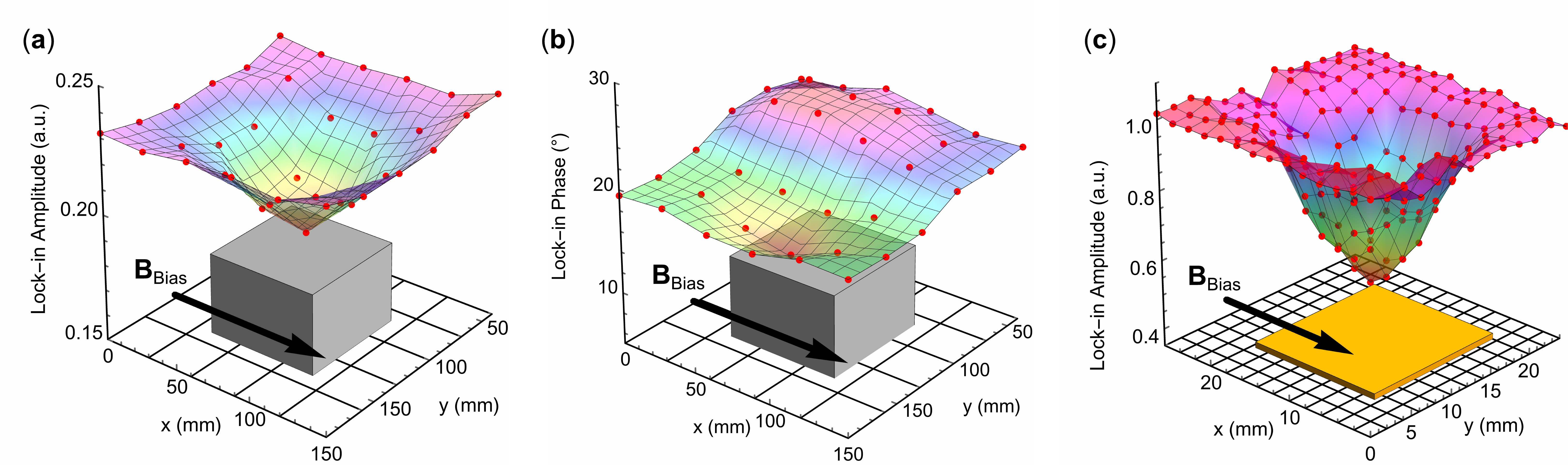}
\caption{The 2-D {conductivity map} of metallic objects as a function of the object position. In (a) and (b) lock-in amplitude and phase signals, respectively, in presence of an Al block $70\times70\times50\, {\rm mm}^3$. It lies $18\,\mathrm{cm}$ above the MOT and $15 \,\mathrm{cm}$ below a $12 \,\mathrm{cm}$ diameter coil. In (c) the 2-D {conductivity map} of a Cu piece  $15\times15\times0.8 \, {\rm mm}^3$. In this case the \textsc{rf} coil was a commercial ferrite coil with a diameter of $4\,\mathrm{mm}$, placed $1\,\mathrm{mm}$ above the object.}

\label{fig:2D_tom}
\end{figure*}


We focused on {the imaging} of non-ferromagnetic, metallic objects (copper, aluminum) by recording the amplitude $A=\sqrt{CHX^2+CHY^2}$ and the phase $\phi=\arctan{(CHY/CHX)}$ of the lock-in signal as a function
of the position of the object. 
We set a constant field $\mathbf{B}_{\rm Bias}$ and a $\textsc{rf}$ radiation in
resonance with the Larmor precession frequency, in order to have a large effect. 
We checked that the presence
of the object didn't change the resonance frequency due to stray magnetic
fields produced by the object itself. Each point {in the image} is the mean over
about 90 measurements, for a typical duration of $18\,\mathrm{s}$ per point.
The object is moved on the $\hat{x}-\hat{y}$ plane with a variable step size and  a precision of  $\pm 0.5\,\mathrm{mm}$.
Given the observed  linear dependence of the lock-in
signals on the atom number, data are normalized to the MOT intensity.
Due to geometric constraints of the setup, the minimum distance of the moving object from the MOT is $18\,\mathrm{cm}$, with the $\textsc{rf}$ coil  placed further vertically at a distance varying from $0.1\,\mathrm{cm}$ to a few centimeters.

A two dimensional {image} of a thick Al cube and of a thin Cu parallelepiped  is shown in
Fig.~\ref{fig:2D_tom}. In the former case both amplitude and phase signals reproduce the shape of the object. The phase is sensitive to the height of walls of the conducting material parallel to the bias magnetic field $\mathbf{B}_{\rm Bias}$, 
as reported in reference~\cite{Bevington2019}. In the Cu case, only the amplitude is reported, as the phase signal is too noisy, probably because this piece is extremely thin. 

\begin{figure}[htb]

\includegraphics[width=0.4\textwidth]{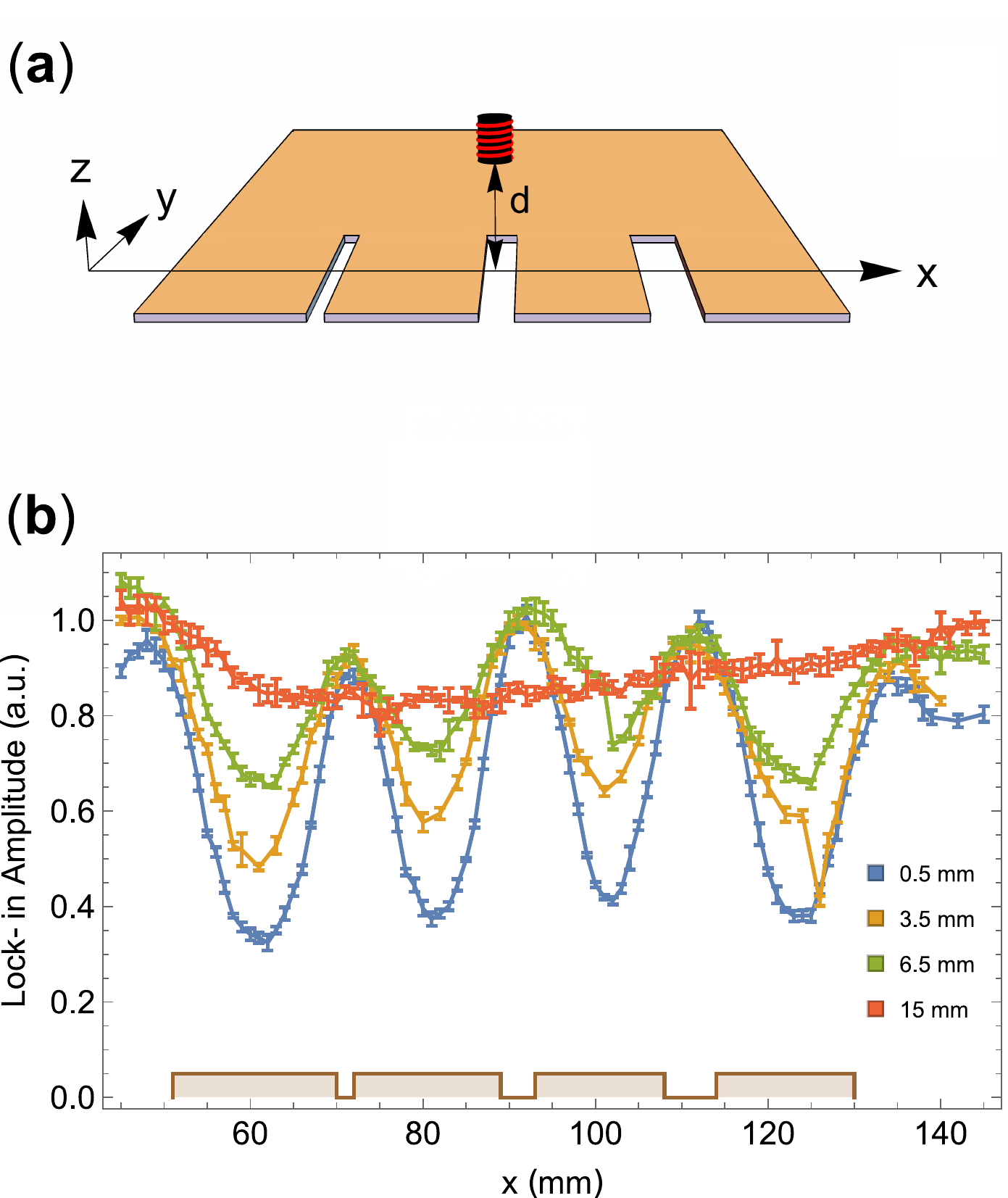}

%

\caption{(a) Scheme of the 1-D {conductivity map} of metallic slabs. (b) Lock-in amplitude vs position $x$ for different coil-slab distances of a Cu parallelepiped with three cuts of 2, 4 and 6$\,\mathrm{mm}$ width.}

\label{fig:1D_tom}
\end{figure}

We investigated the system capability to detect conductivity 
discontinuities.
We studied three evenly spaced cuts of width
2, 4 and 6$\,\mathrm{mm}$, respectively, and
$20\,\mathrm{mm}$ length in one border of a $79\!\times\!79\!\times\!0.8\,\mathrm{mm}$ copper slab as a function of the distance coil/object, with 
 a $6\,\mathrm{mm}$  diameter coil at a
 frequency of $15\,\mathrm{kHz}$. The object was moved in the
direction orthogonal to the cuts
producing the 1-D maps shown in Fig.~\ref{fig:1D_tom}. We found that, at the present level of accuracy, for distances above $15\,\mathrm{mm}$ the detection of these cuts is no more possible. Nevertheless, at $1\,\mathrm{mm}$ detecting distance, also cuts down to $0.4\,\mathrm{mm}$ were mapped. {Assessing the imaging capabilities of our system as a function of the source-object distance is very important in view of applications but is beyond the scope of the present study. We notice that the amplitude of the signal, defined as the difference between the signal in the center of one cut and the value outside the cut, decreases almost linearly in the explored 0.5-15$\,\mathrm{mm}$ interval, but no reliable functional dependence can be extracted so far.}
In order to estimate the spatial resolution of the system, we fitted rising
and falling
profiles of Fig.~\ref{fig:1D_tom} with a linear fit and
calculated the 10 to 90 \% amplitude interval.  We found an average value of about
$5\,\mathrm{mm}$, largely independent of the distance coil-object, similar to the diameter of the \textsc{rf} coil. This is in agreement with ref.~\cite{Bevington2019}, for the regime in which the defect in the conductor is smaller than the coil dimensions.

We now compare the results of our \textsc{rf} C-AM with those of the atomic-vapor OAM, operating in unshielded environment without background image subtraction, of the University College London reported in Ref.~\cite{Deans2016}. While the UCL system provides images with a higher density of pixels, our system shows images with a smaller error (standard deviation of measurements), enhanced contrast in detecting cuts, even with 10 times less averaging, and a slightly better spatial resolution. The latter is mainly limited by the diameter of the \textsc{rf} coils, which is similar in the two cases. The better performances of the cold-atom OAM can be explained in terms of smaller volume of the sensing cold-atom sample with respect to the thermal vapor. On the other hand, the system in Ref.~\cite{Deans2016} operates in a larger range of frequencies and detects objects concealed (although in electric contact) behind a metallic screen, a key requirement for application uses of these sensors. The latter was not possible in our case because C-AM operations in the low-frequency regime ($\leq 1\,\mathrm{kHz}$, where skin depth is larger than a typical screen thickness) were unstable due to background field fluctuations. The implementation of an active field control should solve this issue.

Concerning the footprint of the system, which is also an important requirement for practical applications, we notice that our system is still a ' bulky laboratory machine', occupying roughly 3 cubic meters of space, probably comparable to that in UCL. In literature are nevertheless reported, for instance, portable cold-atom gravity sensors~\cite{Bidel2013} with substantially reduced footprint, or an even smaller cold-atom system, conceived to operate in the International 
Space Station~\cite{Elliott2018}. Similar engineering of the optics and electronics of our setup could be conceived as well, making it more adapted for industrial or security applications.

\label{sec:conclusions}

In conclusion, we demonstrated that a \textsc{rf} C-AM can be used {to obtain conductivity maps} and to detect cuts in metallic objects. The present magnetometer, which operates in an unshielded environment, has a sensitivity of $200 \,\mathrm{pT/\sqrt{Hz}}$ in the $5-35 \,\mathrm{KHz}$ range. It can detect thin ($<1 \,\mathrm{mm}$) objects of about $2 \,\mathrm{cm^{2}}$ surface, as well as cuts down to $0.4 \,\mathrm{mm}$ wide. When the distance between the \textsc{rf} source and the object increases, the resolution {of the map} degrades but larger objects are still clearly detected. The frequency range of the magnetometer can be extended both towards larger as well as smaller frequencies. While the former domain is interesting for medicine and biology~\cite{Deans2020}, the latter is important, for instance, in surveillance and industrial monitoring. 
The present study is limited to the detection of non-magnetic object, thus the obtained images are essentially conductivity maps. Taking amplitude and phase images at different detection frequencies may allow to distinguish between materials with different conductivities and magnetic properties.
We believe that an additional active compensation of stray magnetic fields in the vicinity of the MOT cell~\cite{Bevington2019, Deans2020} would definitely improve the coherence time of the precessing spins and consequently the signal-to-noise ratio.
Finally, the implementation of an optical dipole trap~\cite{Koschorreck2011}, capable of confining the atoms for longer times, should increase the sensitivity, although reducing the repetition rate of the measurements. {In a not-too-far scenario, the use of multiple optical traps or of an optical lattice acting as parallel optical sensors, together with appropriate reconstruction algorithms, could provide a system capable of performing Magnetic Induction Tomography of conductive as well as low-conductivity objects.}

We are indebted with L.~Marmugi and F.~Renzoni for enlightning discussions about the OAM operation, and acknowledge technical assistance from A.~Barbini, F.~Pardini, M.~Tagliaferri and M.~Voliani. This work was partially funded by the ERA-NET Cofund Transnational Call PhotonicSensing—H2020, Grant Agreement
No. 688735, project “Magnetic Induction Tomography with Optical Sensors” (MITOS)/Regione Toscana.

Data of this study are available on request from the authors.


\bibliography{biblio_MIT1}{}
\bibliographystyle{unsrt}

\end{document}